\documentclass[useAMS,usenatbib]{mn2e}

\usepackage{amsmath,array,amsfonts}
\usepackage{amssymb}
\usepackage{graphicx}
\usepackage{natbib}
\usepackage{pstricks}
\usepackage{verbatim}

\newcommand{\ltsima} {$\; \buildrel < \over \sim \;$}
\newcommand{\gtsima} {$\; \buildrel > \over \sim \;$}
\newcommand{\lta} {\lower.5ex\hbox{\ltsima}}
\newcommand{\gta} {\lower.5ex\hbox{\gtsima}}

\title[WMAP 7yr local $f_{\rm NL}$ constraints using neural
  networks]{Constraints on $f_{\rm NL}$ from Wilkinson Anisotropy Probe 7-year data using a neural
  network classifier}

\author[B. Casaponsa et al.]{B. Casaponsa,$^1$ $^2$\thanks{e-mail:
casaponsa@ifca.unican.es} M. Bridges$^3$, A. Curto$^{1}$, R.B. Barreiro$^1$, M.P. Hobson$^3$,\newauthor E.Mart\'{i}nez-Gonz\'{a}lez$^{1}$\\
$^1$     Instituto de F\'isica de Cantabria, CSIC-Universidad de Cantabria, Avda. de los Castros s/n, 39005 Santander, Spain.\\
$^2$     Dpto. de F\'isica Moderna, Universidad de Cantabria, Avda. los Castros s/n, 39005 Santander, Spain.\\
$^3$ Astrophysics Group, Cavendish Laboratory, Madingley Road, Cambridge, CB3 0HE, U.K.}
\date{Accepted ---. Received ---; in original form \today}

\begin{document}

\maketitle

\begin{abstract}
We present a multi-class neural network (NN) classifier as a method to
measure non-Gaussianity, characterised by the local non-linear
coupling parameter $f_{NL}$, in maps of the cosmic microwave
background (CMB) radiation. The classifier is trained on simulated
non-Gaussian CMB maps with a range of known $f_{\rm NL}$ values by
providing it with wavelet coefficients of the maps; we consider both
the {\sc HealPix} (HW) wavelet and the spherical Mexican hat wavelet
(SMHW). When applied to simulated test maps, the NN classfier produces
results in very good agreement with those obtained using standard
$\chi^{2}$ minimization. The standard deviations of the $f_{\rm NL}$
estimates for WMAP-like simulations were $\sigma=22$ and $\sigma=33$
for the SMHW and the HW, respectively, which are extremely close to
those obtained using classical statistical methods in Curto et al. and
Casaponsa et al. Moreover, the NN classifier does not require the
inversion of a large covariance matrix, thus avoiding any need to
regularise the matrix when it is not directly invertible, and is
considerably faster.
\end{abstract}
\section{Introduction}
Artificial intelligence algorithms are being used increasingly to
improve the efficiency of computationally intensive data analysis. In
particular, neural networks (NN) have been successfully applied to
pattern recognition, classification of objects and parameter
estimation in a number of fields, including cosmology \citep[see e.g.][]{Auld2007}.


Cosmological analysis typically involves the use of large datasets and
high precision numerical tools.  In particular, the study of
deviations from Gaussianity in the distribution of temperature
anisotropies in the cosmic microwave background (CMB) require very
demanding computational methods. The simplest way to characterise such
a deviation is through third order moments, as these vanish in the
Gaussian case. It is now commonplace in CMB analysis to work in
spherical harmonic space, where computing the three point correlation
function or bispectrum can prove difficult, or indeed impossible, due
to numerical instability.  Some recent efforts have been applied to
lessen the computational demand without reducing efficiency; see for
example the KSW bispectrum estimator \citep{Komatsu2005}, or the
binned estimator \citep{Bucher2010}.  Other methods which have also
been applied to non-Gaussianity analysis include Minkowski functionals
\citep{Hikage2008,Natoli2010}, wavelet-based methods
\citep{Cayon2001,Mukherjee2004,Curto2009a,Curto2009b,Pietrobon2010,Casaponsa2010},
a Bayesian approach \citep{Elsner2010} and the analysis of the
$N$-dimensional probability density function \citep{Vielva2010}.

This paper introduces an approach based on a neural network classifier
which, after training on third order moments of wavelet coefficients
derived from simulated Gaussian and non-Gaussian CMB realisations, can
be used to estimate the presence and degree of non-Gaussianity for any
given data map.  We have chosen to estimate the local non-linear
coupling parameter $f_{\rm NL}$, which parameterises the local
non-Gaussianity as a quadratic term in the primordial curvature
perturbation.  More precisely, $f_{NL}$ is the amplitude of the
corrections at second order of the primordial curvature perturbations
\citep{Salopek1990,Gangui1994,Verde2000,Komatsu2001}. This type of
non-Gaussianity is predicted even in the simplest slow-roll
inflationary scenario, albeit at a very low level $f_{\rm NL}<~1$,
whereas a wide range of non-standard inflationary models predict much
larger typical $f_{\rm NL}$ values (for a more complete review see
\cite{Bartolo2004},\cite{Babich2004} and \cite{Yadav2010}).
Estimating the value of $f_{\rm NL}$ from a given data map using
existing methods typically has a high computational cost and usually
numerical problems arise (e.g. matrix inversion). As we will show, the
use of neural networks bypasses these difficulties.

In principle, one could use the pixel temperatures in the CMB map
directly, or its spherical harmonic coefficients, as the inputs to the
neural network classifier. Nonetheless, we perform a pre-processing
step in which we decompose the temperature maps into their wavelet
coefficients, which have shown themselves to be sensitive to
non-Gaussian signals
\citep[e.g.][]{Curto2009b,Curto2010,Casaponsa2010}. In particular, we
consider the {\sc HealPix} wavelet (HW) and a spherical Mexican hat
wavelet (SMHW), and compute third-order moments of these wavelet
coefficients, the mean value of which is proportional to $f_{\rm
  NL}$. The network is then trained so that when presented with these
cubic statistics for a new (data) map, it can estimate the $f_{\rm NL}$
value and its error bar. We apply this method to estimate the degree
of non-Gaussianity in the Wilkinson microwave anisotropy probe (WMAP)
7-year data release.

This paper is organized as follows.  In Section~\ref{sec:wavelets}, we
give a brief introduction to the wavelet analysis used. An overview of
the type of neural network employed and our training procedure follows
in Section~\ref{sec:nn}. In Section~\ref{sec:training_data} we explain
the generation of our Gaussian and non-Gaussian simulations, and the
specific characteristics of our $f_{\rm NL}$ classification
network. We present the results of applying our classifier to
simulations and to WMAP 7-year data in Section~\ref{sec:results}. Our
conclusions are summarised in Section~\ref{sec:conclusions}.
  
\section{Wavelets}
\label{sec:wavelets}
Wavelet methods have seen increasing usage in cosmology.  This
has been particularly marked in CMB non-Gaussianity analyses, in which
competitive results have been obtained using wavelets such as the SMHW
\citep{Cayon2003,Vielva2004,Cruz2005,Curto2010}, directional spherical
wavelets \citep{McEwen2008}, spherical Haar wavelet (SHW)
\citep{Tenorio1999,Barreiro2000}, and recently the {\sc HealPix} wavelet
(HW) \citep{Casaponsa2010}. For a review of wavelets applied on the
sphere, see, for example, \cite{McEwen2007}. In essence, decomposing
a CMB map into its wavelet coefficients allows one to separate the
search for non-Gaussianity on different length-scales, while retaining
positional information. In this section we will briefly discuss the
characteristics of both the HW and SMHW and describe how we construct
the statistics which are used in our analysis.
\subsection{ {\sc HealPix} wavelet}
The {\sc HealPix} wavelet is very similar to that presented by
\cite{Shahram2007}. \cite{Casaponsa2010} have used a revised version
of this wavelet and perform the first cosmological application. In
both papers, the central idea is the construction of a fast wavelet
method adapted to the \textsc{HealPix} pixelization scheme
\citep{Gorski2005}. The HW is similar to the SHW in the sense that, at
each level of the wavelet transform, one produces both a high- and
low-resolution map. The low-resolution map for the HW is obtained
simply by averaging over 4-pixel blocks, and the high-resolution map
is just the original map minus the low-resolution map. One begins with
the original map at resolution $J=9$ ($N_\mathrm{side}=512$) and
performs successive wavelet decompositions until resolution $J=2$
($N_{side}=2$), thereby constructing 7 sets of high- and
low-resolution maps. Although the original map is fully represented by
the 7 high-resolution maps plus the low-resolution map at $J=2$, in
our analysis we have used all the high- and low-resolution maps, plus
the original map, since this has been shown to improve results
\citep[see][for details]{Casaponsa2010}. 

Using all these maps, the third order moments of the wavelet
coefficients are computed as follows:
\begin{equation}
 S_{jkl}=\frac{1}{N_{l}}\frac{\sum_{i=1}^{N_{l}}w_{i,j}w_{i,k}w_{i,l}}{\sigma_{j}\sigma_{k}\sigma_{l}},
\label{eq:stat}
\end{equation}
where $w_{i,j}$ is the $i^{\rm th}$ wavelet coefficient of the map at
resolution $j$, $\sigma_{j}$ is the dispersion of $w_{i,j}$, and
$N_{l}$ is the number of pixels in the map at resolution $l$ (since
one requires $j \leq k \leq l$).  Some of these statistics are
redundant (linearly dependency exists between them), so we restrict
our analysis to the set of non-redudant statistics, which gives a
total of 232 quantities; these are then computed for non-Gaussian
simulations with a range of known values of $f_{\rm NL}$.

The expected values of these statistics are proportional to the
non-linear coupling parameter, and they have previously been used to
estimate the best fit $f_{\rm NL}$ value for the data by weighted
least squares parameter estimation \citep{Casaponsa2010}. In this case, the
dispersion in the estimated $f_{\rm NL}$ value for Gaussian
simulations and is found to be $\sigma(f_{\rm NL})=34$, which is
slightly larger that the optimal value. The main
advantage of the HW is the computing efficiency; for example, the
third-order statistics construction is $10^{3}$ times faster than for
the KSW bispectrum estimator \citep{Komatsu2005} and $10^{2}$ times
faster than the SMHW (see below). This procedure (for both the HW and
SMHW) does, however, include the estimation and inversion of a
correlation matrix, which can be computationally demanding and, in
some cases, close to singular. As we will show below, this step is
avoided with the use of a NN classifier.

\subsection{Spherical Mexican Hat Wavelet }
The spherical Mexican hat wavelet (SMHW)
\citep{antoine1998,martinez2002} has produced competitive results in
constraining primordial non-Gaussianity
\citep{Mukherjee2004,Curto2009a,Curto2009b,Curto2010}. It is a
continuous wavelet that has better frequency localization than the HW,
although the computing efficiency is lower.  \cite{Curto2010} use the
SMHW to constrain $f_{\rm NL}$ with an accuracy equivalent to the
bispectrum estimators \citep[see for example][]{Smith2009,fergusson2009,fergusson2010,Komatsu2010,Bucher2010}.
The definition of the third-order moments is the same as for the
HW. In this case, however, there are more inter-scale combinations
because the scales involved are not restricted by the \textsc{HealPix}
pixelization. The total number of non-redundant statistics for the
SMHW wavelet coefficients is 680. Using the mean values and
covariances of these statistics computed from non-Gaussian
simulations, \cite{Curto2010} applied a $\chi^{2}$ minimisation method
to obtain optimal uncertainties on the $f_{\rm NL}$ estimates of
$\sigma \approx 21$. However, this method requires a principal
component analysis to deal with the degenerancies present in the
covariance matrix. As we will see, this problem is avoided with the
use of the multi-class neural network classifier.

\section{Neural networks}
\label{sec:nn}
Artificial neural networks are a methodology for computing, based on
massive parallelism and redundancy, which are features also found in
animal brains. They consist of a number of interconnected processors
each of which processes information and passes it to other processors
in the network. Well-designed networks are able to `learn' from a set
of training data and to make predictions when presented with new,
possibly incomplete, data.  These algorithms have been successfully
applied in several areas, in particular, we note the following
applications in astrophysics:
\cite{Storrie-Lombardi1992,Baccigalupi2000,Vanzella2004,Auld2007} and
\cite{Carballo2008}.
 
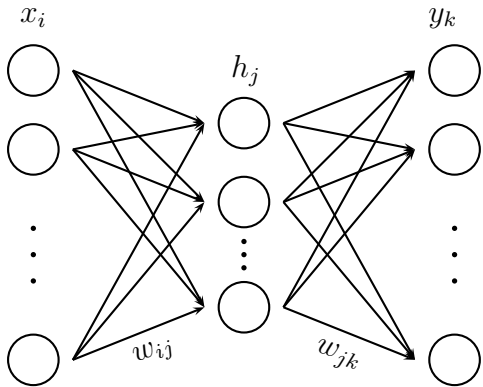
\begin{figure}
 \center
\psset{unit=0.7cm}
\begin{pspicture}(0,0)(10,8)
\uput[0](0.5,7){\Large$x_{i}$}
\pscircle(1,6){0.5}
\pscircle(1,4.5){0.5}
\psdot[dotsize=2pt](1,3)
\psdot[dotsize=2pt](1,2.5)
\psdot[dotsize=2pt](1,2)
\pscircle(1,0.5){0.5}
\rput[bl]{20}(3,0.3){\Large$w_{ij}$}
\psline{->}(1.75,6)(4.25,5)
\psline{->}(1.75,6)(4.25,3.5)
\psline{->}(1.75,6)(4.25,1.5)
\psline{->}(1.75,4.5)(4.25,5)
\psline{->}(1.75,4.5)(4.25,3.5)
\psline{->}(1.75,4.5)(4.25,1.5)
\psline{->}(1.75,0.5)(4.25,5)
\psline{->}(1.75,0.5)(4.25,3.5)
\psline{->}(1.75,0.5)(4.25,1.5)
\uput[0](4.5,6){\Large$h_{j}$}
\pscircle(5,5){0.5}
\pscircle(5,3.5){0.5}
\psdot[dotsize=2pt](5,2.75)
\psdot[dotsize=2pt](5,2.5)
\psdot[dotsize=2pt](5,2.25)
\pscircle(5,1.5){0.5}
\rput[bl]{-18}(6.3,0.5){\Large$w_{jk}$}
\psline{->}(5.75,5)(8.25,6)
\psline{->}(5.75,5)(8.25,4.5)
\psline{->}(5.75,5)(8.25,0.5)
\psline{->}(5.75,3.5)(8.25,6)
\psline{->}(5.75,3.5)(8.25,4.5)
\psline{->}(5.75,3.5)(8.25,0.5)
\psline{->}(5.75,1.5)(8.25,6)
\psline{->}(5.75,1.5)(8.25,4.5)
\psline{->}(5.75,1.5)(8.25,0.5)
\uput[0](8.25,7){\Large$y_{k}$}
\pscircle(9,6){0.5}
\pscircle(9,4.5){0.5}
\psdot[dotsize=2pt](9,3)
\psdot[dotsize=2pt](9,2.5)
\psdot[dotsize=2pt](9,2)
\pscircle(9,0.5){0.5}

\end{pspicture}
\caption{Schematic of a 3-layer feed-forward neural network.}
\label{fig:network_diagram}
 \end{figure}

The basic building block of an ANN is the \emph{neuron}. Information
is passed as inputs to the neuron, which processes them and produces
an output.  The output is typically a simple mathematical function of
the inputs. The power of the ANN comes from assembling many neurons
into a network. The network is able to model very complex behaviour
from input to output. We use a three-layer feed-forward network
consisting of a layer of input neurons, a layer of `hidden' neurons
and a layer of output neurons.  In such an arrangement each neuron is
referred to as a node. Figure~\ref{fig:network_diagram} shows a
schematic design of such a network.

The outputs of the hidden layer and the output layer are related to
their inputs as follows:
\begin{eqnarray}
\mbox{hidden layer:} & h_j=g^{(1)}(f_j^{(1)}); &
f_j^{(1)} = \sum_i w^{(1)}_{ji}x_i +
  b_j^{(1)}, \\
\mbox{output layer:} & y_k=g^{(2)}(f_k^{(2)}); & f_k^{(2)} =
\sum_j w^{(2)}_{kj}h_j + b_k^{(2)},\label{eq:outputs}
\end{eqnarray}
where the output of the hidden layer $h$ and output layer $y$ are
given for each hidden node $j$ and each output node $k$. The index $i$
runs over all input nodes. The functions $g^{(1)}$ and $g^{(2)}$ are
called activation functions. The non-linear nature of $g^{(1)}$ is a
key ingredient in constructing a viable and practically useful
network. This non-linear function must be bounded, smooth and
monotonic; we use $g^{(1)}(x) = \tanh x$. For $g^{(2)}$ we simply use
$g^{(2)}(x) = x$.  The layout and number of nodes are collectively
termed the \emph{architecture} of the network. For a basic introduction 
to artificial neural networks the reader is directed to \citet{MacKay}.

For a given architecture, the weights $\mathbf{w}$ and biases
$\mathbf{b}$ define the operation of the network and are the
quantities we wish to determine by some \emph{training} algorithm. We
denote $\mathbf{w}$ and $\mathbf{b}$ collectively by $\mathbf{a}$. As
these parameters vary during training, a very wide range of non-linear
mappings between inputs and outputs is possible. In fact, according to
a `universal approximation theorem'~\cite{Leshno1993}, a standard
three-layer feed-forward network can approximate any continuous
function to \emph{any} degree of accuracy with appropriately chosen
activation functions and a sufficient number of hidden nodes. 
 
In our application, we will construct a {\em classification} network.
The aim of any classification method is to place members of a set into
groups based on inherent properties or \emph{features} of the
individuals, given some pre-classified training data.  Formally,
classification can be summarised as finding a classifier $\mathcal{C}
: \mathbf{x} \rightarrow C$ which maps an object from some (typically
multi-dimensional) feature space $\mathbf{x}$ to its classification
label $C$, which is typically taken as one of $ \left\{1, ...,N
\right\}$ where $N$ is the number of distinct classes.  Thus the
problem of classification is to partition feature space into regions
(not necessarily contiguous), assigning each region a label
corresponding to the appropriate classification.  In our context, the
aim is to classify sets of third-order statistics of wavelet
coefficients of (possibly) non-Gaussian CMB maps (assembled into an
input feature vector $\mathbf{x}$) into classes defined by ranges of
$f_{\rm NL}$; this is discussed in more detail below.

In building a classifier using a neural network, it is convenient to
view the problem \emph{probabilistically}. To this end we consider a
3-layer MLP (multi-layer percepton) consisting of an input layer
($x_i$), a hidden layer ($h_j$), and an output layer ($y_i$).  In
classification networks, however, the outputs are transformed
according to the \emph{softmax} procedure
\begin{equation}
p_k = \frac{e^{y_k}}{\sum_m e^{y_m}},
\end{equation}
such that they are all non-negative and sum to unity. In this way
$p_k$ can be interpreted as the probability that the input feature
vector $\mathbf{x}$ belongs to the $k$th class. A suitable objective
function for the classification problem is then
\begin{equation}
\mathcal{L}(\mathbf{a}) = \sum_{l} \sum_{k}
t^{(l)}_k \ln p_k(\mathbf{x}^{(l)}, \mathbf{a}),
\end{equation}
where the index $l$ runs over the training dataset $\mathcal{D} =
\{\mathbf{x}^{(l)},\mathbf{t}^{(l)}\}$, in which the target vector
$\mathbf{t}^{(l)}$ for the network outputs has unity in the element
corresponding to the true class of the $l^{\rm th}$ feature vector
$\mathbf{x}^{(l)}$ and zeroes elsewhere.  One then wishes to choose
network parameters $\mathbf{a}$ so as to maximise this objective
function as the training progresses.  The advantage of this
probabilistic approach is that we gain the ability to make
\emph{statistical} decisions on the appropriate classification in very
large feature spaces where a direct linear partition would not be
feasible.

One wishes to choose network parameters $\mathbf{a}$ so as to maximise
the objective function $\mathcal{L}(\mathbf{a})$ as the training
progresses. This is, however, a highly non-linear, multi-modal
function in many dimensions whose optimisation poses a non-trivial
problem. We perform this optimisation using the {\sc MemSys} package
\citep{Gull1999}. This algorithm considers the parameters $\mathbf{a}$ 
to have prior probabilities proportional to $e^{\alpha S(\mathbf{a})}$, 
where $S(\mathbf{a})$ is the positive-negative entropy functional
\citep{Hobson1998}. $\alpha$ is treated as a hyper-parameter of the prior,
and sets the scale over which variations in $\mathbf{a}$ are 
expected. $\alpha$ is chosen to maximise its marginal posterior probability 
whose value is inversely proportional to the standard deviation of the prior.
Thus for a given $\alpha$, the log-posterior probability is proportional
to $\mathcal{L}(\mathbf{a}) + \alpha S(\mathbf{a})$. For each chosen
$\alpha$ there is a solution $\hat{\mathbf{a}}$ that maximises the posterior.
As $\alpha$ varies, the set of solutions $\hat{\mathbf{a}}$ is called the 
\emph{maximum-entropy trajectory}. We wish to find the solution for which
$\mathcal{L}$ is maximised which occurs at the end of the trajectory where $\alpha=0$.
For practical purposes we start at a large value of $\alpha$ and iterate
downwards until $\alpha$ is sufficiently small so that the posterior
is dominated by the $\mathcal{L}$ term. {\sc MemSys} performs this algorithm 
using conjugate gradient descent at each step to converge to the maximum-entropy
trajectory. The required matrix of second derivatives of $\mathcal{L}$ is 
approximated using vector routines only, thus circumventing the need for
$O(N^3)$ operations required for exact calculations. The application of {\sc MemSys} to the problem of
network training allows for the fast efficient training of relatively
large network structures on large data sets that would otherwise be
difficult to perform in a reasonable time. Moreover the {\sc MemSys}
package also computes the Bayesian evidence for the model
(i.e. network) under consideration, \cite[see for example][for a
 review]{Jaynes2003}, which provides a powerful model selection
tool. In principle, values of the evidence computed for each possible
architecture of the network (and training data) provide a mechanism to
select the most appropriate architecture, which is simply the one that
maximises the evidence (although we will use a more prosaic method
below for deciding on the network architecture). The {\sc MemSys}
algorithm is described in greater detail in \citep{Gull1999}.

\section{The $f_{\rm NL}$ classification network}
\label{sec:training_data}

To train our $f_{\rm NL}$ classification network we provide it with an
ensemble of training data $\mathcal{D} =
\{\mathbf{x}^{(l)},\mathbf{t}^{(l)}\}$. The $l^{\rm th}$ input vector
$\mathbf{x}^{(l)}$ contains the third-order statistics of the wavelet
coefficients of the $l^{\rm th}$ simulated CMB map.  The output classes of
our network correspond to contiguous ranges of $f_{\rm NL}$
values. Thus, the target vector $\mathbf{t}^{(l)}$ for the network
outputs has zeroes everywhere except for a unit entry in the element
corresponding to the class in which the true $f_{\rm NL}$ value of the
$l^{\rm th}$ simulated CMB map falls.

The $N$ output classes of the network were defined by dividing some
initial (anticipated) range of $f_{\rm NL}$ values into $N$
equal-width subranges. For example, for a total range of $-30\le
f_{\rm NL}<30$ and a network with just 3 output classes, input vectors
constructed from maps with $-30\le f_{\rm NL}<-10$ were ascribed to
\textit{class}=1 with an associated target vector
$\mathbf{t}=(1,0,0)$, maps with $-10\le f_{\rm NL}<10$ to
\textit{class}=2 with $\mathbf{t}=(0,1,0)$, and those with $10\le
f_{NL}<30$ to \textit{class}=3 with $\mathbf{t}=(0,0,1)$. In this
example, the output given by the network for some test input vector
$\mathbf{x}$ would be a 3-dimensional vector
$\mathbf{p}=(p_{1},p_{2},p_{3})$, where $\sum_k p_{k}=1$ and $p_{k}$
can be interpreted as the probability that the input vector belongs to
class $k$.  The discrepancy between the targets and the outputs during
training can be measured by the true positive rate, which is simply
the fraction of the training input vectors for which the network
assigns the maximum probability to the correct class.

From the output values $p_k$ obtained for each map, we define the
estimator of the local non-Gaussianity parameter to be simply
\begin{equation}
\hat{f}_{\rm NL}=\sum_{k=1}^{n_\mathrm{class}} \langle f_{\rm NL}
\rangle_k p_{k}
\label{eq:fnl_est}
\end{equation}
where $\langle f_{\rm NL} \rangle_k$ is the mean value of $f_{\rm NL}$
in the $k^{\rm th}$ class. The statistical properties of this estimator,
namely its mean and dispersion, determine the accuracy of the method.

\subsection{Training data}

The training input vectors $\mathbf{x}^{(l)}$ were generated as
follows.  We began with a set of 1000 non-Gaussian CMB realisations
from which $a_{lm}^{\rm NG}$ and $a_{lm}^{\rm G}$ were generated by
\cite{Elsner2009} and normalized to the WMAP 7-year concordance model
power spectrum generated by CAMB.  These $a_{lm}$ are publicly
available\footnote{http://planck.mpa-garching.mpg.de/cmb/fnl-simulations/}.
The ultimate accuracy of the network classifier is improved, however,
by the inclusion of further training data. Given the finite number of
available simulations, we thus created a further set by rotation of
the original maps by $90^{\circ}$ perpendicular to the galactic
plane. This rotation creates roughly 20 per cent extra map area based
on the original mask; we verified that its inclusion improves the
results. Using this procedure we generate a further 1000 non-Gaussian
simulations. Of the 2000 non-Gaussian maps, 1800 were used for
training and the remainder were set aside for testing of the networks.

For each non-Gaussian simulation used for training, sets of
$a_{lm}$ were then generated with varying $f_{NL}$ using
the following prescription
 \begin{equation}
a_{lm}=a_{lm}^{\rm G}+f_{\rm NL} a_{lm}^{\rm NG},
\end{equation}
with 20 different $f_{\rm NL}$ random values between $-120$
and $120$ for the HW decomposition and between $-76$ and $76$ for the SMHW analysis. Thus, for each non-Gaussian simulation, 20 sets of
$a_{lm}$ were generated. Hence the total number of available training
data sets is 36000. Noise-weighted V+W-band WMAP realizations were
then constructed as explained in \cite{Curto2009a} and
\cite{Casaponsa2010}, and the KQ75 mask was then applied, which
covers roughly $29\%$ of the sky. A wavelet decomposition for both the
HW and SMHW was performed to determine the wavelet coefficents for
each $a_{lm}$ set, and their third-order moments computed.  These
statistics were provided as inputs to the neural network. Each input
vector contained 232 values for the HW and 680 for the SMHW.

\subsection{Network architecture}
\label{subsec:net_arch}
The architecture of our 3-layer neural networks are defined by two
free parameters: the number of hidden nodes $n_\mathrm{hid}$ and the
number of output classes, $n_\mathrm{class}$, into which the $f_{\rm
  NL}$ range is divided.  A further parameter, which determines the
accuracy of the network classifier, is the quantity of training data
$n_\mathrm{data}$.  Variation of these parameters can affect the
training efficiency so it is desirable to explore this training space
adequately in order to find an optimal set of parameters.

Although the {\sc MemSys} algorithm provides routines to determine the
optimal value of the number of hidden nodes using the Bayesian
evidence \cite{Gull1999}, in this application $n_{\rm hid}$ is
determined simply by measuring training times and the accuracy of the
trained networks on an independent testing set. In this example, we
have found that in fact the optimal architecture contains no hidden
nodes, resulting in what is effectively a linear classifier. This is
not surprising, since we are effectively `asking' the network to learn
the mean value and dispersion of the third-order moments of the
wavelet coefficients for each $f_{NL}$; since the expectation value is
linearly dependent on the $f_{NL}$, this network architecture
trivially satisfies this requirment. Indeed, networks of this sort
provide a simple way of obtaining the (pseudo)inverse of any matrix.

The number of output classes, $n_\mathrm{class}$, of the network is
clearly related to the total range of $f_{\rm NL}$ considered and size
of the subranges into which this range is divided.  Here we consider
networks with $n_\mathrm{class}=9$ (an odd number ensures that $f_{\rm
  NL}=0$ does not lie on the boundary of a class) The range of $f_{\rm
  NL}$ was chosen \emph{a priori} to correspond to approximately
$\pm3\sigma$, where $\sigma$ is the dispersion in the $f_{\rm NL}$
estimates obtained previously using the standard $\chi^2$ minimisation
method. Thus, the full range was taken to be $-120 \le f_{\rm NL} <
120$ for the HW and $-76 \le f_{\rm NL}< 76$ for the SMHW, resulting
in subranges per class of width 27 and 17 units, respectively. This
combination fulfilled all the requirements of classification over the
range of our simulated data.

\begin{figure}
\centering
\resizebox{7cm}{!}{\includegraphics{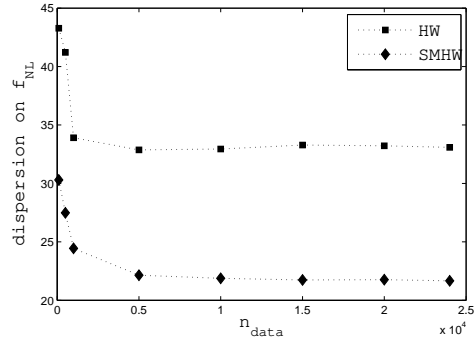}} 
\caption{Results of the dispersion of $\hat{f}_{\rm NL}$ for 1000 Gaussian
  simulations for different values of $n_\mathrm{data}$.}
\label{fig:sigma_ndata}
\end{figure}

The quantity of training data, $n_{\rm data}$, determines the accuracy
of the resulting classification network. Naturally, the network
accuracy increases with $n_{\rm data}$, but it typically saturates
after a given number. We found that the quantity of data
required saturated at roughly $n_\mathrm{data}\sim10000$ (see
Fig.~\ref{fig:sigma_ndata}).  

\subsection{Training evolution}

\begin{figure*}
\centering
\includegraphics[width=12cm,height=8cm]{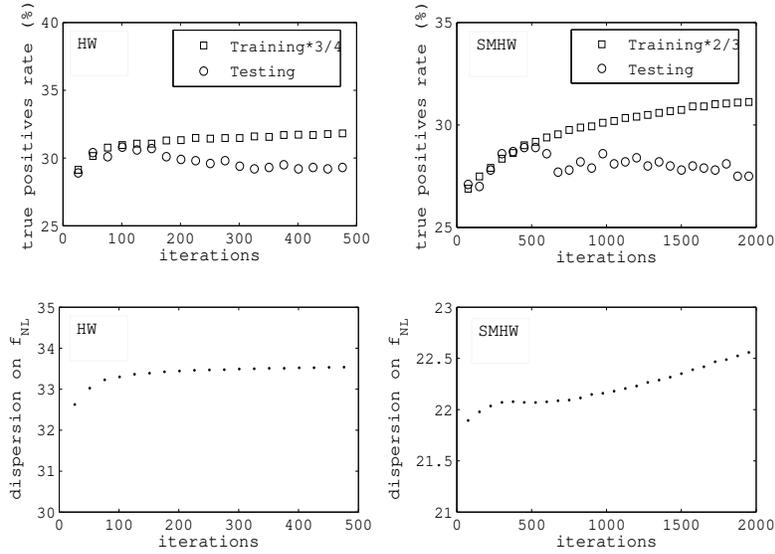} 
\caption{Evolution of the true positive rate for each iteration of the
  training process with a neural network with $n_\mathrm{hid}=0$ and
  $n_\mathrm{data}=10000$. Note that the TPR of the training set have
  been multiplied by a factor less than unity in order to highlight
  the divergence of the behaviours. The bottom panels show the
  variation of the dispersion on the estimate $\hat{f}_{\rm NL}$
  during the training. Left panels for HW and right panels for SMHW.}
\label{fig:true_positives}
\end{figure*}

Figure \ref{fig:true_positives} illustrates the training evolution for
the classification network with $n_{\rm hid}=0$ and $n_{\rm
  class}=9$. In the top two panels we plot the true positive rates
(TPR) of the network on the training set and the test set, for the HW
and SHMW respectively; in each plot, the TPR on the training set has
been mutliplied by a factor less than unity to highlight the
divergence with the TPR for the test set. We see that this divergence
occurs after $\sim 100$ and $\sim 500$ iterations of the {\sc MemSys}
optimiser for the HW and SMHW, respectively. Thus the training was
terminated at this point to construct our final classification
networks.

A key criterion in determining the quality of our classifiers is the
dispersion of the $f_{\rm NL}$ values obtained in the testing set.
This is plotted in the bottom two panels of Figure
\ref{fig:true_positives} for the HW and SMHW, respectively.  We note
that, in each case, this dispersion increases noticeable beyond the
point where the TPRs on the training and testing sets diverge.

\section{Results}
\label{sec:results}
\subsection{Application to WMAP simulations}

\begin{figure}
\begin{center}$
\begin{array}{c}
\includegraphics[width=5cm,height=4cm]{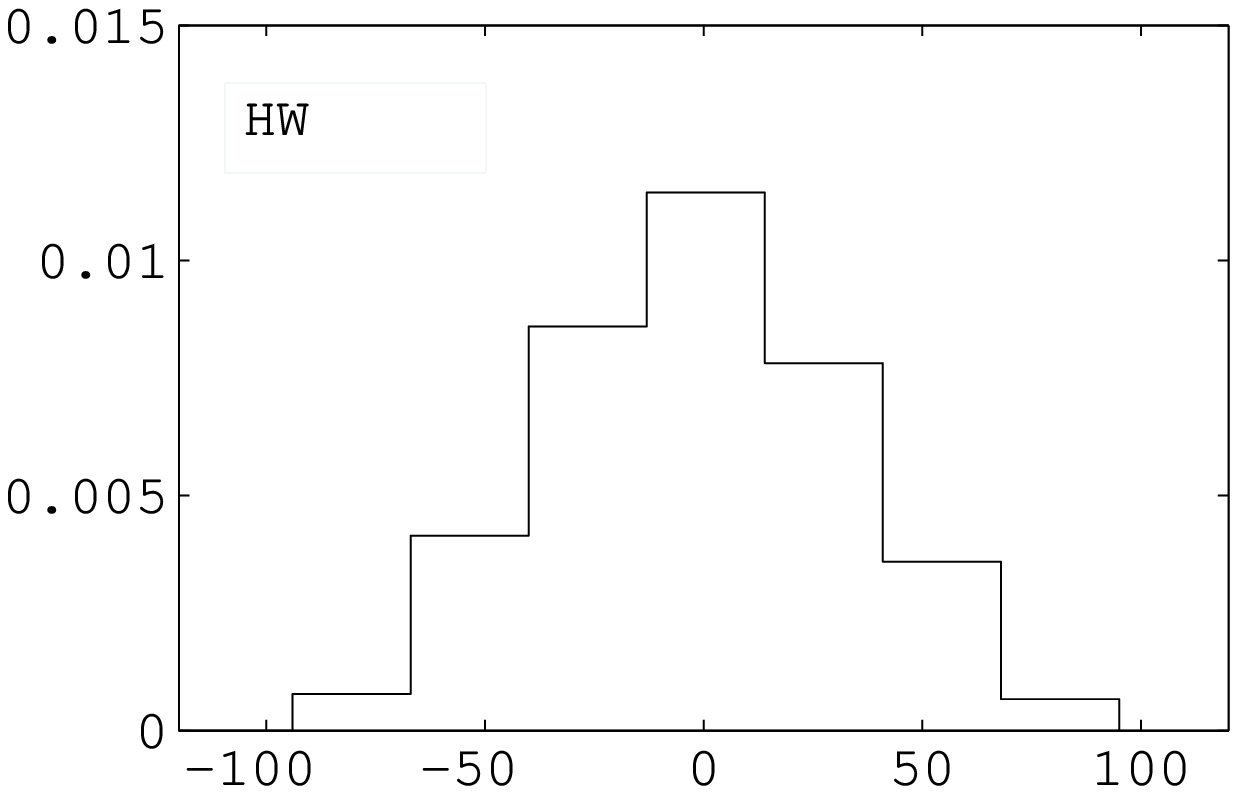} \\ 
\includegraphics[width=5cm,height=4cm]{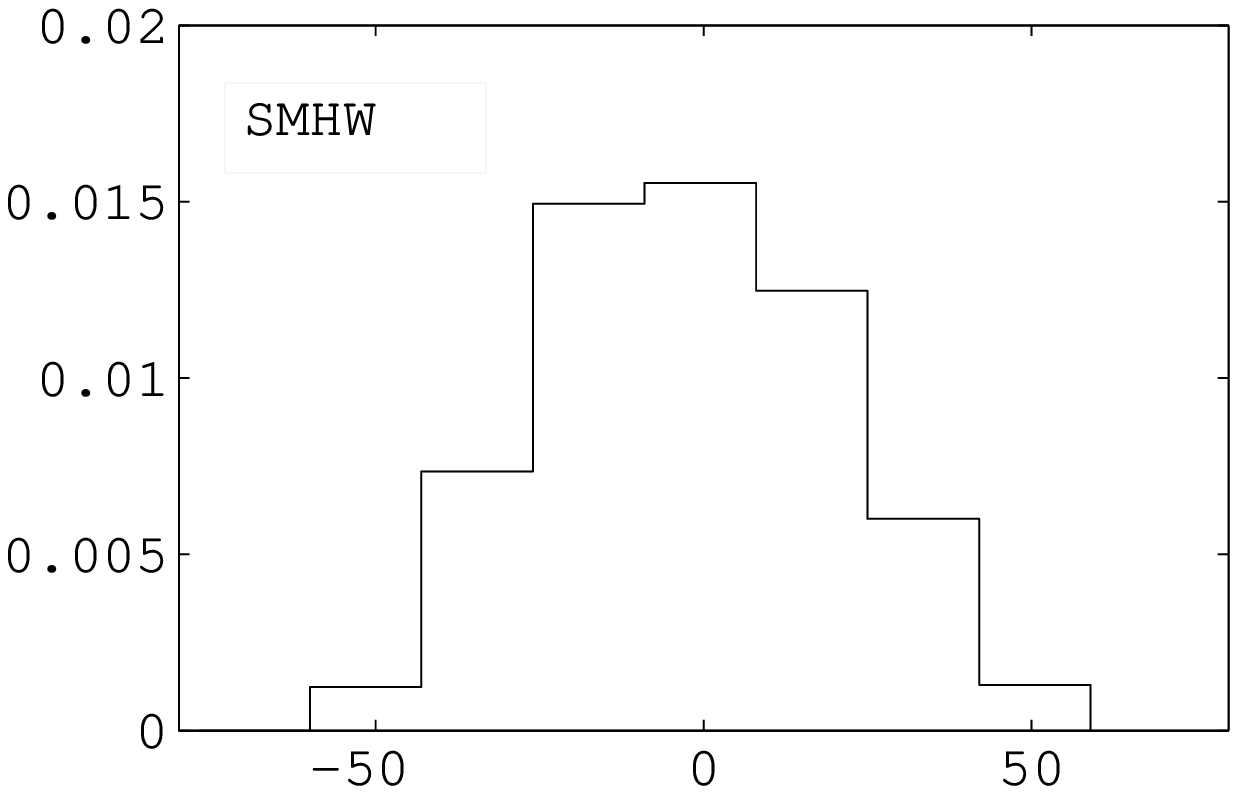} 
\end{array}$
\end{center}
\caption{Distribution of $\hat{f}_{\rm NL}$ obtained from 1000
  Gaussian realizations for HW (top) and SMHW (bottom).}
\label{fig:hist}
 \end{figure}

We first applied our classifiers to 1000 WMAP-7yr simulations made
from Gaussian CMB maps ($f_{\rm NL}=0$). For the HW classifier, we
obtained $\langle\hat{f}_{\rm NL}\rangle=-1$, which indicates the
estimator is essentially unbiassed. Moreover, the dispersion of the
estimator $\sigma(\hat{f}_{\rm NL})=33$ is extremely similar to that
obtained with the weighted least-squares method
($\sigma(\hat{f}_{NL})=34$).  The full distribution of the estimator
is shown in the top panel of Fig.~\ref{fig:hist}. For the SMHW
classifier, we again found $\langle\hat{f}_{\rm NL}\rangle=-1$, with a
dispersion of $\sigma(\hat{f}_{\rm NL})=22$, which is very close to
the optimal value of $\sigma(\hat{f}_{\rm NL})=21$.  The distribution
of the estimator for the SMHW is shown in the bottom panel of
Fig.~\ref{fig:hist}.

The histogram bins in Fig.~\ref{fig:hist} have the same size and
central values as those used to define the network classes.  We see
that the classes at extremal $f_{\rm NL}$ values are empty, indicating
that the network placed no maps in these $f_{\rm NL}$ ranges. Thus
for estimating $f_{\rm NL}$ from Gaussian or nearly Gaussian maps the
range in $f_{\rm NL}$ used is sufficiently wide.

\begin{figure}
\begin{center}$
\begin{array}{c}
\includegraphics[width=6cm,height=4cm]{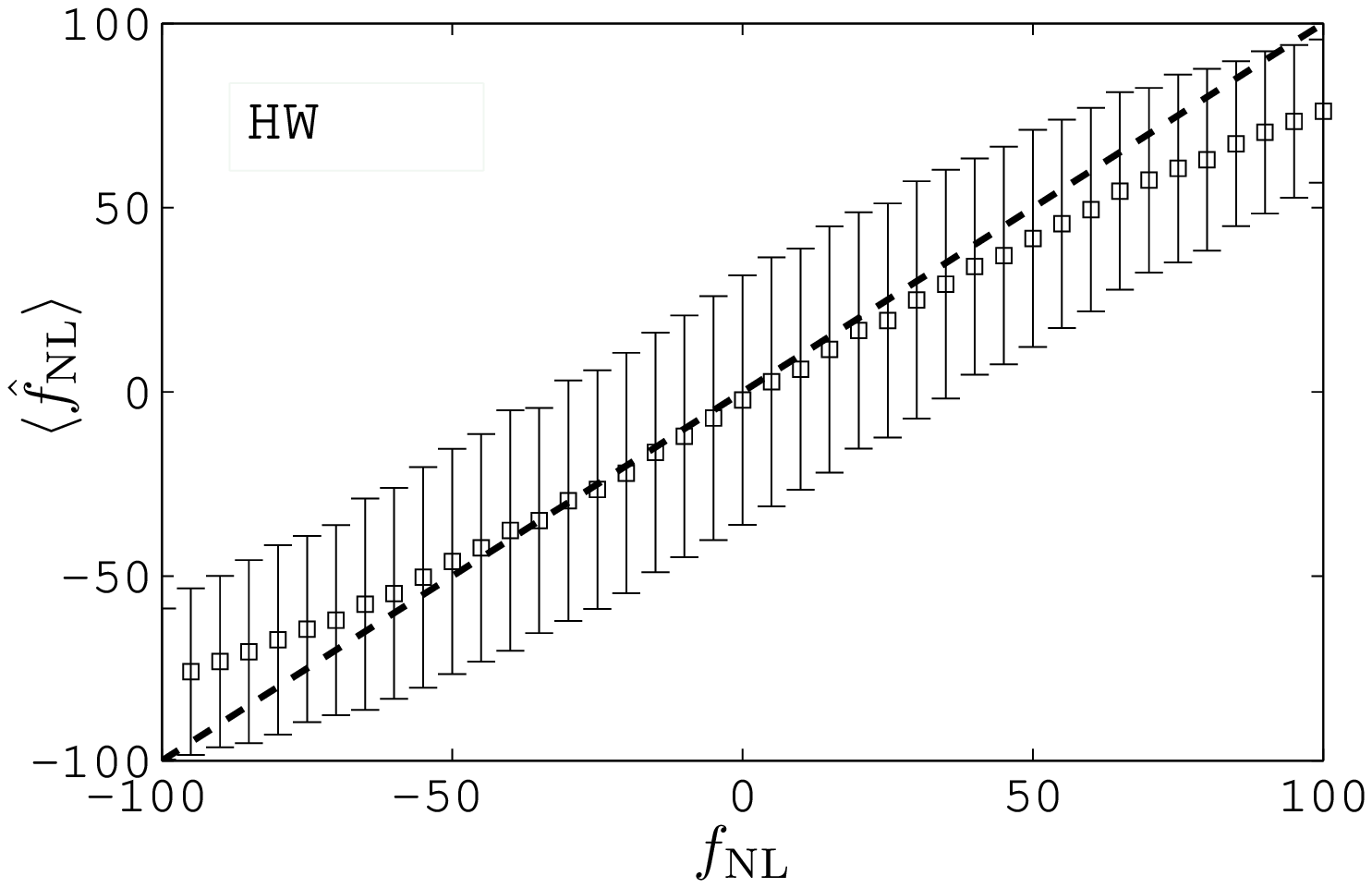} \\ 
\includegraphics[width=6cm,height=4cm]{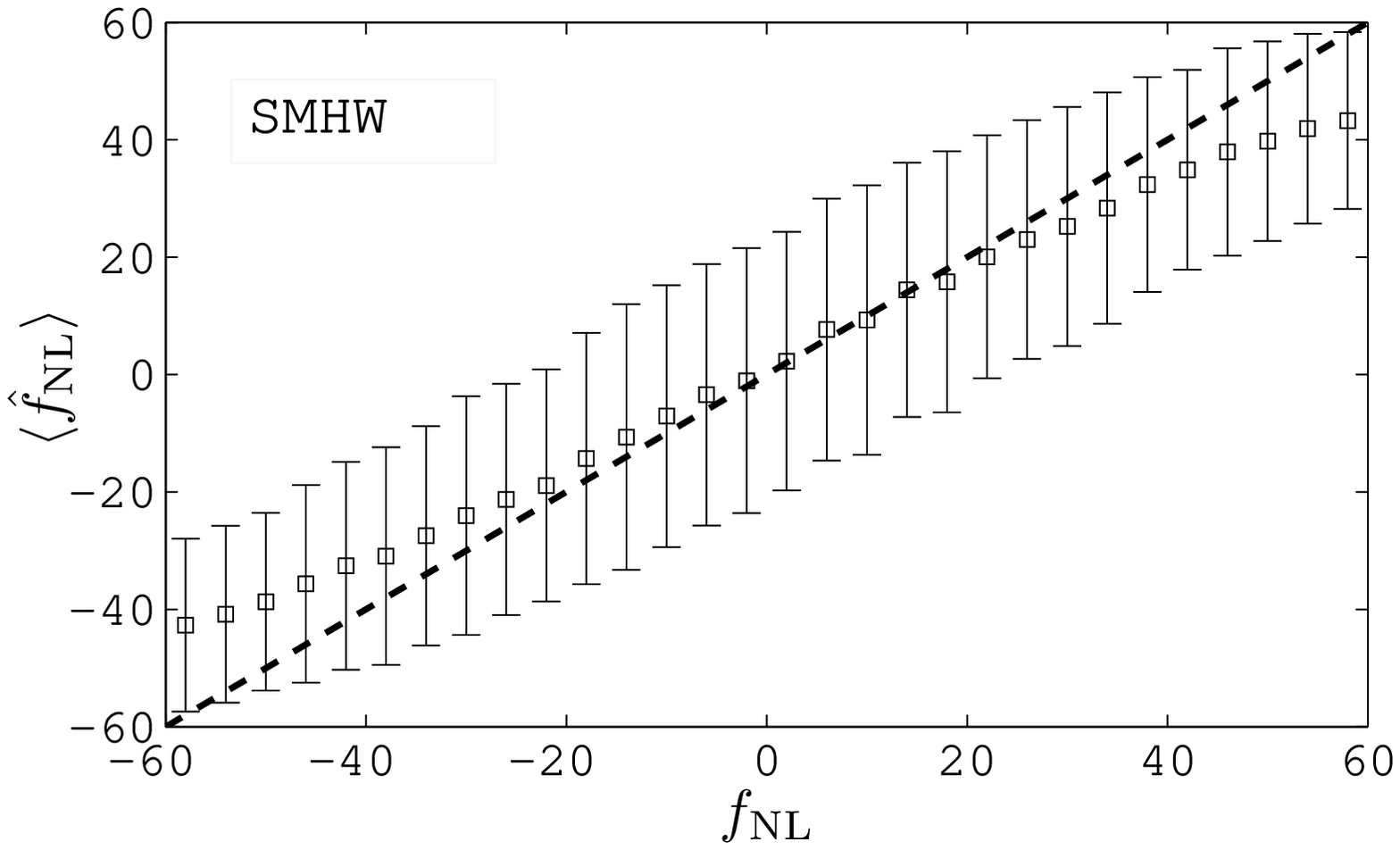}
\end{array}$
\end{center}
\caption{The mean and dispersion of $\hat{f}_{\rm NL}$ obtained for a
  number of representative values of the true $f_{\rm NL}$ for the HW
  network (top) and the SMHW network (bottom).} 
\label{fig:diff_fnl}
\end{figure}

We next applied our estimator to sets of non-Gaussian simulations,
each with a different non-zero $f_{\rm NL}$ value. For each true
$f_{\rm NL}$ value, we analysed the corresponding WMAP simulations and
calculated the mean and dispersion of our estimator $\hat{f}_{\rm NL}$
for both the HW and SMHW classifiers. The results are shown in
fig.~\ref{fig:diff_fnl}, in which we plot the mean value of
$\hat{f}_{\rm NL}$ against the true $f_{\rm NL}$ value. We see that
the classifiers are unbiassed for $|f_{\rm NL}|\lesssim \sigma$ with
an almost constant dispersion.  For larger $|f_{\rm NL}|$ values,
however, the estimator becomes progressively more biassed and its
dispersion decreases.

\begin{figure*}
\begin{center}$
\begin{array}{cc} 
\includegraphics[width=9cm,height=7cm]{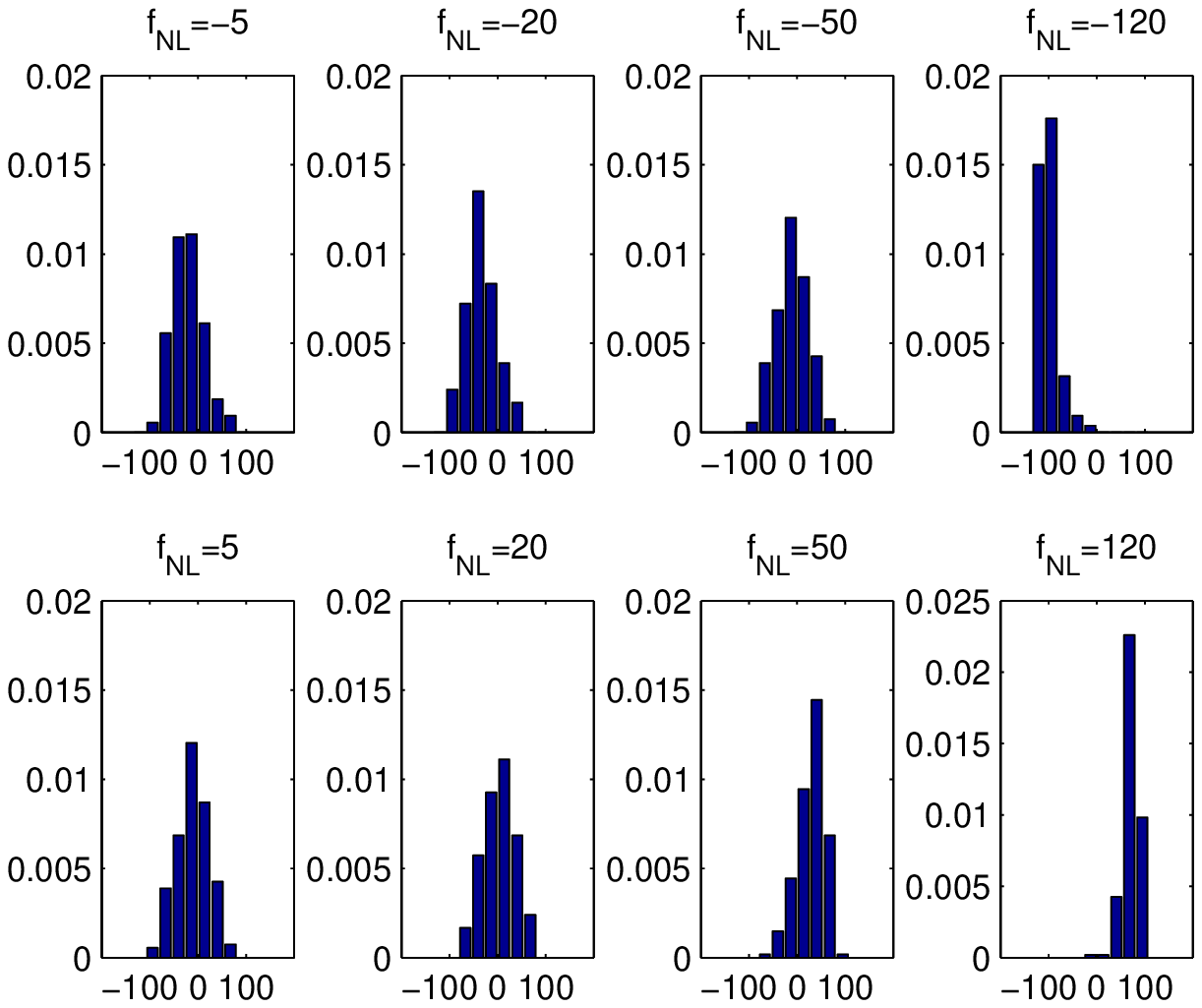}
\includegraphics[width=9cm,height=7cm]{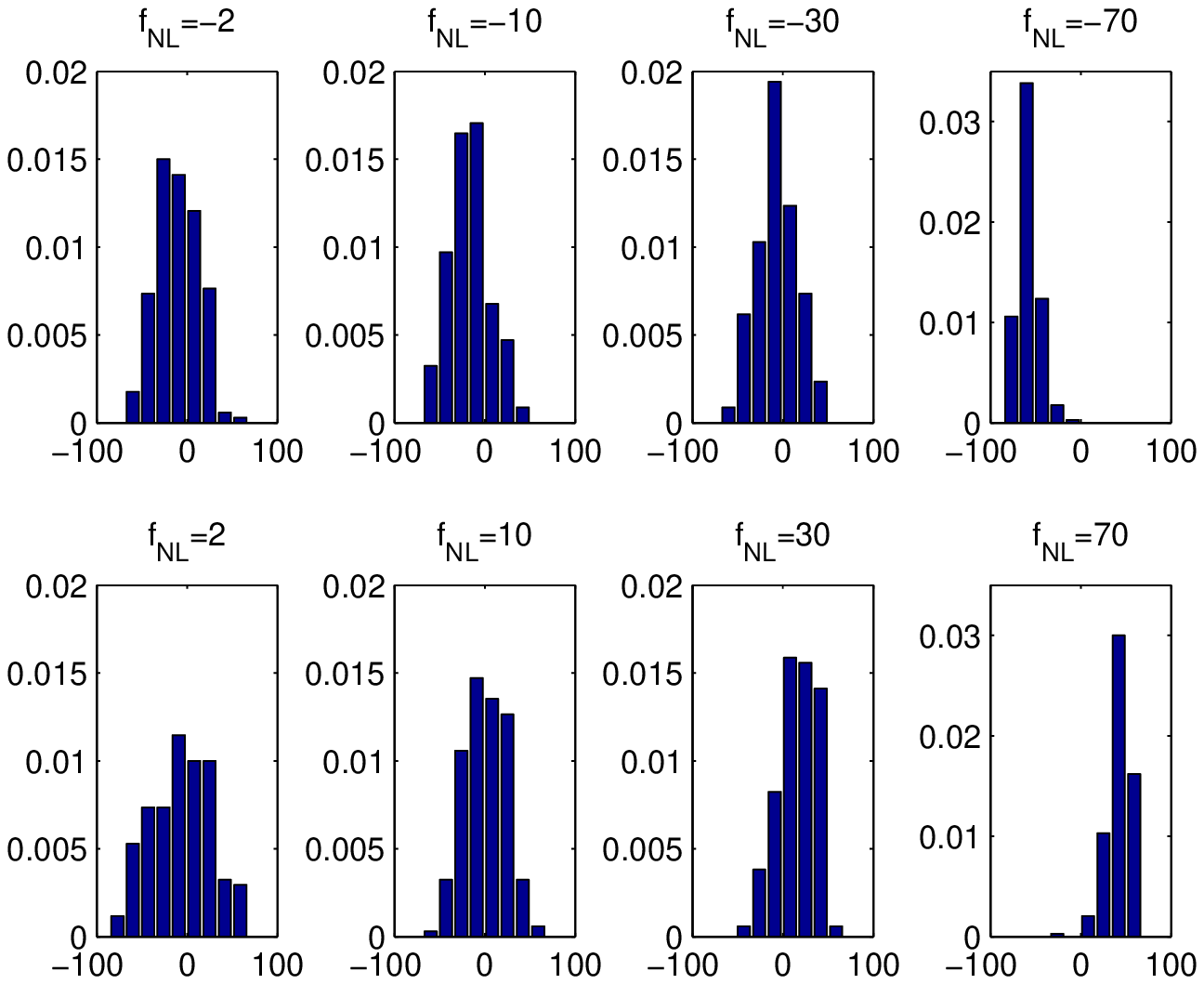} 
\end{array}$
\end{center}
\caption{Distribution of $\hat{f}_{\rm NL}$ obtained from 200
  non-Gaussian realizations with representative true $f_{\rm NL}$
  values, for HW (left) and SMHW (right).}
\label{fig:histos_fnl}
\end{figure*}

The latter behaviour is simply understood as an \emph{edge effect} due
to the finite total range of $f_{\rm NL}$ considered by the networks.
This point is illustrated in Fig.~\ref{fig:histos_fnl} in which we
plot the full distributions of $\hat{f}_{\rm NL}$ obtained for a
number of representative values of the true $f_{\rm NL}$. We see that
for $|f_{\rm NL}|\lesssim \sigma$, we obtain close to symmetric
distribution centred on the true $f_{\rm NL}$ value, with no maps
being placed in the extreme classes. As $|f_{\rm NL}|> \sigma$,
however, we see that the classifier does begin to place maps in the
extreme classes, resulting in the distribution of $\hat{f}_{\rm NL}$
becoming severely skewed and no longer centred on the true value.
Of course, if one were to encounter this behaviour in the analysis of
a real data set, one could simply alter the range of $f_{\rm NL}$
considered by the network and retrain.

In any case, the above results show that both the HW and SMHW network
classifiers produce unbiassed estimates $\hat{f}_{\rm NL}$ provided
$-\sigma<f_{NL}<\sigma$. Moreover, the dispersions on these estimators
are very similar to those obtained with the classical weighted least
squares (WLS) method, indicating that neural networks can produce very
accurate results within the limitations described above.  In the case
of the SMHW, this is a particularly important result since the
complexity of the covariance matrix inversion required in the standard
approach is bypassed via the use of the neural network classifier.
\cite{Curto2010} used a principal component analysis to reduce the
covariance matrix dimension to allow inversion.
\newpage
\begin{table*}
\centering
\begin{tabular}{l r r r r r}
\hline\hline
& $\hat{f}^{\phantom{\int}}_{\rm NL,data}$ &$\sigma(\hat{f}_{\rm NL})$ &$\langle\hat{f}_{\rm NL,gauss}\rangle$ & $P_{2.5}$&$P_{97.5}$\\[0.5ex]\hline\hline\\
SMHW (NN) &19 & 22&$-1$ & $-43$& 42\\\\
SMHW (WLS) & & & & & \\[-1ex]
\tiny{Curto et al. 2011b}& \raisebox{1.5ex} {37} & \raisebox{1.5ex}{21} & \raisebox{1.5ex} {0}& \raisebox{1.5ex}{$-42$} &\raisebox{1.5ex}{46}\\\\
HW (NN) & $-12$ & 33  & $-1$ &$-66$ & 63 \\\\
HW (WLS)& & & & &\\[-1ex] \tiny{Casaponsa et al. 2011}& \raisebox{1.5ex}{6} &\raisebox{1.5ex}{34} & \raisebox{1.5ex}{1}&\raisebox{1.5ex}{$-68$} & \raisebox{1.5ex}{67} \\\\\hline
\end{tabular}
\caption{Results obtained with neural networks (NN)
  and weighted least squares (WLS). $\hat{f}_{\rm NL,data}$ is the best
  fitting value for V+W WMAP data, $\langle\hat{f}_{\rm NL,gauss}\rangle$ and
  $\sigma(\hat{f}_{\rm NL})$ are the expected value and the standard
  deviation for Gaussian simulations. $P_{2.5}$ and
  $P_{97.5}$ represent the percentile values at 95\% confidence level
  of $\hat{f}_{\rm NL}$ for Gaussian
  realizations.} \label{tab:results}
\end{table*}
\subsection{Application to WMAP 7-year data}

Applying the neural network classifiers to real data (the V+W WMAP
7-year data map), we obtain $\hat{f}_{\rm NL}=-12$ for the HW and
$\hat{f}_{\rm NL}=19$ for the SMHW. Both these values lie well within
the corresponding dispersion of the estimator. From the corresponding
$\hat{f}_{\rm NL}$ distributions obtained on simulated data, we find
that 95\% confidence limits are $-78<f_{\rm NL}<51$ for the HW and
$-24<f_{\rm NL}<61$ for the SMHW.\footnote{Note that the constraints
  are not corrected for the unresolved point sources contribution.}
We therefore conclude that the data are consistent with the Gaussian
hypothesis. We note that the SMHW confidence limits are very similar
to those obtained with the optimal $f_{\rm NL}$ estimator
\citep{Komatsu2010,Smith2009}.

These results are summarised in Table~\ref{tab:results}, along with
found via the weighted least squares (WLS) method. The latter results
are also consistent with Gaussianity. It is worth mentioning, however,
the different values of $\hat{f}_{\rm NL}$ obtained by the neural
network and the WLS methods, for both HW and SMHW. Although all four
values lie well within their corresponding dispersions, each method
returns a different $\hat{f}_{\rm NL}$ value when applied to the same
WMAP-7yr dataset. This behaviour is to be expected, however, since these
are four {\em different} estimators of $f_{\rm NL}$. Therefore, in
general, they will not be equal, even when applied to the same input
data. Only the statistical properties (e.g. mean, dispersion) of their
sampling distributions are important.

\section{Conclusions}
\label{sec:conclusions} 
We have trained a multi-class neural network classifier with
third-order moments of the HW and SMHW coefficients of non-Gaussian
realizations in order to set constraints on the local non-linear
coupling term $f_{NL}$ using WMAP 7-year data.  We found that with a
very simple network and with few iterations (requiring just a few secs
CPU time) it is possible to produce the same results as those obtained
with the weighted least squares method.  This is an interesting
achievement, as it bypasses any covariance matrix related computations
and their associated regularisation problems.  The estimation of the
covariance matrix with both wavelets requires the analysis of at least
10000 Gaussian simulations which involves a huge demand in CPU time,
in particular with the SMHW statisitcs.  The error bars on the
estimation of $f_{\rm NL}$ computed with Gaussian simulations are
$\sigma(\hat{f}_{NL}=33)$ for HW and $\sigma(\hat{f}_{NL})=22$ for
SMHW, which are extremely similar to the ones obtained in
\cite{Casaponsa2010} and \cite{Curto2010} using the same statisitcs
but a different estimator based on the weighted least squares method
($\sigma=34$, $\sigma=21$ for HW and SMHW respectively). The
constraints for WMAP 7-year data were found to be $-78<f_{\rm NL}<51$
for the HW and $-24<f_{\rm NL}<61$ for the SMHW, which are compatible
to a Gaussian distribution as found in
\cite{Smith2009,Curto2009b,Komatsu2010,Casaponsa2010} and Curto et
al. (2011b). The results obtained with the SMHW
statistics are similar to the ones found in \cite{Smith2009} and
\cite{Komatsu2010}, which are the most stringent ones currently
available at the limit of the WMAP resolution.  Further analysis, as
to the contribution to $f_{\rm NL}$ of unresolved point sources or
foregrounds can be performed by applying the linear classifier to the
statistics of new simulated maps with this characteristic signal.

\section*{acknowledgments}
We acknowledge partial financial support from the Spanish Ministerio
de Ciencia e Innovaci\'on project 
AYA2010-21766-C03-01 and from the
CSIC-The Royal Society joint project with reference 2008GB0012. 
B. Casaponsa thanks the Spanish Ministerio de
Ciencia e Innovaci\'on for a pre-doctoral fellowship. The
authors acknowledge the computer resources, technical expertise and
assistance provided by the Spanish Supercomputing Network (RES) node
at Universidad de Cantabria. We acknowledge the use of Legacy Archive
for Microwave Background Data Analysis (LAMBDA). Support for it is
provided by the NASA Office of Space Science. The {\sc HealPix} package was
used throughout the data analysis \citep{Gorski2005}.
\bibliographystyle{mn2e}
\bibliography{nn-version6}

\begin{thebibliography}{}

\bibitem[\protect\citeauthoryear{{Antoine} \& {Vandergheynst}}{{Antoine} \&
  {Vandergheynst}}{1998}]{antoine1998}
{Antoine} J.-P.,  {Vandergheynst} P.,  1998, Journal of Mathematical Physics,
  39, 3987

\bibitem[\protect\citeauthoryear{{Auld}, {Bridges}, {Hobson} \& {Gull}}{{Auld}
  et~al.}{2007}]{Auld2007}
{Auld} T.,  {Bridges} M.,  {Hobson} M.~P.,    {Gull} S.~F.,  2007, \mnras, 376,
  L11

\bibitem[\protect\citeauthoryear{{Babich}, {Creminelli} \&
  {Zaldarriaga}}{{Babich} et~al.}{2004}]{Babich2004}
{Babich} D.,  {Creminelli} P.,    {Zaldarriaga} M.,  2004, Journal of Cosmology
  and Astro-Particle Physics, 8, 9

\bibitem[\protect\citeauthoryear{{Baccigalupi}, {Bedini}, {Burigana}, {De
  Zotti}, {Farusi}, {Maino}, {Maris}, {Perrotta}, {Salerno}, {Toffolatti} \&
  {Tonazzini}}{{Baccigalupi} et~al.}{2000}]{Baccigalupi2000}
{Baccigalupi} C.,  {Bedini} L.,  {Burigana} C.,  {De Zotti} G.,  {Farusi} A.,
  {Maino} D.,  {Maris} M.,  {Perrotta} F.,  {Salerno} E.,  {Toffolatti} L.,
  {Tonazzini} A.,  2000, \mnras, 318, 769

\bibitem[\protect\citeauthoryear{{Barreiro}, {Hobson}, {Lasenby}, {Banday},
  {G{\'o}rski} \& {Hinshaw}}{{Barreiro} et~al.}{2000}]{Barreiro2000}
{Barreiro} R.~B.,  {Hobson} M.~P.,  {Lasenby} A.~N.,  {Banday} A.~J.,
  {G{\'o}rski} K.~M.,    {Hinshaw} G.,  2000, \mnras, 318, 475

\bibitem[\protect\citeauthoryear{{Bartolo}, {Komatsu}, {Matarrese} \&
  {Riotto}}{{Bartolo} et~al.}{2004}]{Bartolo2004}
{Bartolo} N.,  {Komatsu} E.,  {Matarrese} S.,    {Riotto} A.,  2004, \prd, 402,
  103

\bibitem[\protect\citeauthoryear{{Bucher}, {van Tent} \& {Carvalho}}{{Bucher}
  et~al.}{2010}]{Bucher2010}
{Bucher} M.,  {van Tent} B.,    {Carvalho} C.~S.,  2010, \mnras, 407, 2193

\bibitem[\protect\citeauthoryear{{Carballo}, {Gonz{\'a}lez-Serrano}, {Benn} \&
  {Jim{\'e}nez-Luj{\'a}n}}{{Carballo} et~al.}{2008}]{Carballo2008}
{Carballo} R.,  {Gonz{\'a}lez-Serrano} J.~I.,  {Benn} C.~R.,
  {Jim{\'e}nez-Luj{\'a}n} F.,  2008, \mnras, 391, 369

\bibitem[\protect\citeauthoryear{{Casaponsa}, {Barreiro}, {Curto},
  {Mart{\'{\i}}nez-Gonz{\'a}lez} \& {Vielva}}{{Casaponsa}
  et~al.}{2010}]{Casaponsa2010}
{Casaponsa} B.,  {Barreiro} R.~B.,  {Curto} A.,  {Mart{\'{\i}}nez-Gonz{\'a}lez}
  E.,    {Vielva} P.,  2010, ArXiv e-prints

\bibitem[\protect\citeauthoryear{{Cay{\'o}n}, {Mart{\'{\i}}nez-Gonz{\'a}lez},
  {Arg{\"u}eso}, {Banday} \& {G{\'o}rski}}{{Cay{\'o}n}
  et~al.}{2003}]{Cayon2003}
{Cay{\'o}n} L.,  {Mart{\'{\i}}nez-Gonz{\'a}lez} E.,  {Arg{\"u}eso} F.,
  {Banday} A.~J.,    {G{\'o}rski} K.~M.,  2003, \mnras, 339, 1189

\bibitem[\protect\citeauthoryear{{Cay\'{o}n}, {Sanz}, {Martinez-Gonzalez},
  {Banday}, {Argueso}, {Gallegos}, {Gorski} \& {Hinshaw}}{{Cay\'{o}n}
  et~al.}{2001}]{Cayon2001}
{Cay\'{o}n} L.,  {Sanz} J.~L.,  {Martinez-Gonzalez} E.,  {Banday} A.~J.,
  {Argueso} F.,  {Gallegos} J.~E.,  {Gorski} K.~M.,    {Hinshaw} G.,  2001,
  \mnras, 326, 1243

\bibitem[\protect\citeauthoryear{{Cruz}, {Mart{\'{\i}}nez-Gonz{\'a}lez},
  {Vielva} \& {Cay{\'o}n}}{{Cruz} et~al.}{2005}]{Cruz2005}
{Cruz} M.,  {Mart{\'{\i}}nez-Gonz{\'a}lez} E.,  {Vielva} P.,    {Cay{\'o}n} L.,
   2005, \mnras, 356, 29

\bibitem[\protect\citeauthoryear{{Curto}, {Mart{\'{\i}}nez-Gonz{\'a}lez} \&
  {Barreiro}}{{Curto} et~al.}{2009b}]{Curto2009b}
{Curto} A.,  {Mart{\'{\i}}nez-Gonz{\'a}lez} E.,    {Barreiro} R.~B.,  2009b,
  \apj, 706, 399

 \bibitem[\protect\citeauthoryear{{Curto}, {Martinez-Gonzalez} \&
  {Barreiro}}{{Curto} et~al.}{2011a}]{Curto2010}
{Curto} A.,  {Martinez-Gonzalez} E.,    {Barreiro} R.~B.,  2011a, \mnras, 412, 1023

\bibitem[\protect\citeauthoryear{{Curto}, {Martinez-Gonzalez}, 
  {Barreiro}\& {Hobson} }{{Curto} et~al.}{2011b}]{Curto2011}
{Curto} A.,  {Martinez-Gonzalez} E.,    {Barreiro} R.~B.,  {Hobson} M.~P., 2011b, submitted to \mnras

\bibitem[\protect\citeauthoryear{{Curto}, {Mart{\'{\i}}nez-Gonz{\'a}lez},
  {Mukherjee}, {Barreiro}, {Hansen}, {Liguori} \& {Matarrese}}{{Curto}
  et~al.}{2009a}]{Curto2009a}
{Curto} A.,  {Mart{\'{\i}}nez-Gonz{\'a}lez} E.,  {Mukherjee} P.,  {Barreiro}
  R.~B.,  {Hansen} F.~K.,  {Liguori} M.,    {Matarrese} S.,  2009a, \mnras, 393,
  615

\bibitem[\protect\citeauthoryear{{Elsner} \& {Wandelt}}{{Elsner} \&
  {Wandelt}}{2009}]{Elsner2009}
{Elsner} F.,  {Wandelt} B.~D.,  2009, \apjs, 184, 264

\bibitem[\protect\citeauthoryear{{Elsner} \& {Wandelt}}{{Elsner} \&
  {Wandelt}}{2010}]{Elsner2010}
{Elsner} F.,  {Wandelt} B.~D.,  2010, \apj, 724, 1262

\bibitem[\protect\citeauthoryear{{Fergusson}, {Liguori} \&
  {Shellard}}{{Fergusson} et~al.}{2010}]{fergusson2010}
{Fergusson} J.~R.,  {Liguori} M.,    {Shellard} E.~P.~S.,  2010, Physical
  Review D, 82, 023502

\bibitem[\protect\citeauthoryear{{Fergusson} \& {Shellard}}{{Fergusson} \&
  {Shellard}}{2009}]{fergusson2009}
{Fergusson} J.~R.,  {Shellard} E.~P.~S.,  2009, Phys. Rev. D, 80, 043510

\bibitem[\protect\citeauthoryear{{Gangui}, {Lucchin}, {Matarrese} \&
  {Mollerach}}{{Gangui} et~al.}{1994}]{Gangui1994}
{Gangui} A.,  {Lucchin} F.,  {Matarrese} S.,    {Mollerach} S.,  1994, \apj,
  430, 447

\bibitem[\protect\citeauthoryear{{G{\'o}rski}, {Hivon}, {Banday}, {Wandelt},
  {Hansen}, {Reinecke} \& {Bartelmann}}{{G{\'o}rski} et~al.}{2005}]{Gorski2005}
{G{\'o}rski} K.~M.,  {Hivon} E.,  {Banday} A.~J.,  {Wandelt} B.~D.,  {Hansen}
  F.~K.,  {Reinecke} M.,    {Bartelmann} M.,  2005, \apj, 622, 759

\bibitem[\protect\citeauthoryear{{Gull} \& {Skilling}}{{Gull} \&
  {Skilling}}{1999}]{Gull1999}
{Gull} S.~F.,  {Skilling} J.,  1999, {Quantified maximum entropy: Mem-Sys 5
  users’ manual}.
Maximum Entropy Data Consultants Ltd, Royston

\bibitem[\protect\citeauthoryear{{Hikage}, {Matsubara}, {Coles}, {Liguori},
  {Hansen} \& {Matarrese}}{{Hikage} et~al.}{2008}]{Hikage2008}
{Hikage} C.,  {Matsubara} T.,  {Coles} P.,  {Liguori} M.,  {Hansen} F.~K.,
  {Matarrese} S.,  2008, \mnras, 389, 1439

\bibitem[\protect\citeauthoryear{{Hobson} \& {Lasenby}}{{Hobson} \&
  {Lasenby}}{1998}]{Hobson1998}
{Hobson} M.,  {Lasenby} A.,  1998, 298, 905

\bibitem[\protect\citeauthoryear{{Jaynes}}{{Jaynes}}{2003}]{Jaynes2003}
{Jaynes} E.,  2003, {Probability Theory: The Logic of Science}.
Cambridge University Press

\bibitem[\protect\citeauthoryear{{Komatsu}, {Smith}, {Dunkley}, {Bennett},
  {Gold}, {Hinshaw}, {Jarosik}, {Larson}, {Nolta}, {Page}, {Spergel} \&
  {Halpern} M. an~{Wright}}{{Komatsu} et~al.}{2011}]{Komatsu2010}
{Komatsu} E.,  {Smith} K.~M.,  {Dunkley} J.,  {Bennett} C.~L.,  {Gold} B.,
  {Hinshaw} G.,  {Jarosik} N.,  {Larson} D.,  {Nolta} M.~R.,  {Page} L.,
  {Spergel} D.~N.,    {Halpern} M. an~{Wright} E.~L.,  2011, \apjs, 192, 18

\bibitem[\protect\citeauthoryear{{Komatsu} \& {Spergel}}{{Komatsu} \&
  {Spergel}}{2001}]{Komatsu2001}
{Komatsu} E.,  {Spergel} D.,  2001, \prd, 63, 063002

\bibitem[\protect\citeauthoryear{{Komatsu}, {Spergel} \& {Wandelt}}{{Komatsu}
  et~al.}{2005}]{Komatsu2005}
{Komatsu} E.,  {Spergel} D.~N.,    {Wandelt} B.~D.,  2005, \apj, 634, 14

\bibitem[\protect\citeauthoryear{{Leshno}, V., A. \& {Schocken}}{{Leshno}
  et~al.}{1993}]{Leshno1993}
{Leshno} M.,  V. Y.,  A. P.,    {Schocken} S.,  1993, Neural Netw., 6, 861

\bibitem[\protect\citeauthoryear{MacKay}{MacKay}{2003}]{MacKay}
MacKay D.,  2003, {Information Theory, Inference and Learning Algorithms}.
Cambridge University Press

\bibitem[\protect\citeauthoryear{{Mart\'{i}nez-Gonz\'{a}lez}, {Gallegos},
  {Argueso}, {Cayon} \& {Sanz}}{{Mart\'{i}nez-Gonz\'{a}lez}
  et~al.}{2002}]{martinez2002}
{Mart\'{i}nez-Gonz\'{a}lez} E.,  {Gallegos} J.~E.,  {Argueso} F.,  {Cayon} L.,
    {Sanz} J.~L.,  2002, \mnras, 336, 22

\bibitem[\protect\citeauthoryear{{McEwen}, {Hobson}, {Lasenby} \&
  {Mortlock}}{{McEwen} et~al.}{2008}]{McEwen2008}
{McEwen} J.~D.,  {Hobson} M.~P.,  {Lasenby} A.~N.,    {Mortlock} D.~J.,  2008,
  \mnras, 388, 659

\bibitem[\protect\citeauthoryear{{McEwen}, {Vielva}, {Wiaux}, {Barreiro},
  {Cayon}, {Hobson}, {Lasenby}, {Martinez-Gonzalez} \& {Sanz}}{{McEwen}
  et~al.}{2007}]{McEwen2007}
{McEwen} J.~D.,  {Vielva} P.,  {Wiaux} Y.,  {Barreiro} R.~B.,  {Cayon} L.,
  {Hobson} M.~P.,  {Lasenby} A.~N.,  {Martinez-Gonzalez} E.,    {Sanz} J.~L.,
  2007, Journal of Fourier Analysis and Applications, 13, 495

\bibitem[\protect\citeauthoryear{{Mukherjee} \& {Wang}}{{Mukherjee} \&
  {Wang}}{2004}]{Mukherjee2004}
{Mukherjee} P.,  {Wang} Y.,  2004, \apj, 613, 51

\bibitem[\protect\citeauthoryear{{Natoli}, {de Troia}, {Hikage}, {Komatsu},
  {Migliaccio}, {Ade}, {Bock}, {Bond}, {Borrill}, {Boscaleri}, {Contaldi},
  {Crill}, {de Bernardis}, {de Gasperis}, {de Oliveira-Costa}, {di Stefano} \&
  {Hivon}}{{Natoli} et~al.}{2010}]{Natoli2010}
{Natoli} P.,  {de Troia} G.,  {Hikage} C.,  {Komatsu} E.,  {Migliaccio} M.,
  {Ade} P.~A.~R.,  {Bock} J.~J.,  {Bond} J.~R.,  {Borrill} J.,  {Boscaleri} A.,
   {Contaldi} C.~R.,  {Crill} B.~P.,  {de Bernardis} P.,  {de Gasperis} G.,
  {de Oliveira-Costa} A.,  {di Stefano} G.,    {Hivon} E.,  2010, \mnras, 408,
  1658

\bibitem[\protect\citeauthoryear{{Pietrobon}}{{Pietrobon}}{2010}]{Pietrobon201%
0}
{Pietrobon} D.,  2010, Memorie della Societa Astronomica Italiana Supplementi,
  14, 278

\bibitem[\protect\citeauthoryear{{Salopek} \& {Bond}}{{Salopek} \&
  {Bond}}{1990}]{Salopek1990}
{Salopek} D.~S.,  {Bond} J.~R.,  1990, \prd, 42, 3936

\bibitem[\protect\citeauthoryear{{Shahram}, {Donoho} \& {Starck}}{{Shahram}
  et~al.}{2007}]{Shahram2007}
{Shahram} M.,  {Donoho} D.,    {Starck} J.,  2007, in Society of Photo-Optical
  Instrumentation Engineers (SPIE) Conference Series Vol.~6701 of Society of
  Photo-Optical Instrumentation Engineers (SPIE) Conference Series, {Multiscale
  representation for data on the sphere and applications to geopotential data}

\bibitem[\protect\citeauthoryear{{Smith}, {Senatore} \& {Zaldarriaga}}{{Smith}
  et~al.}{2009}]{Smith2009}
{Smith} K.~M.,  {Senatore} L.,    {Zaldarriaga} M.,  2009, Journal of Cosmology
  and Astro-Particle Physics, 9, 6

\bibitem[\protect\citeauthoryear{{Storrie-Lombardi}, {Lahav}, {Sodre} Jr. \&
  {Storrie-Lombardi}}{{Storrie-Lombardi} et~al.}{1992}]{Storrie-Lombardi1992}
{Storrie-Lombardi} M.~C.,  {Lahav} O.,  {Sodre} Jr. L.,    {Storrie-Lombardi}
  L.~J.,  1992, \mnras, 259, 8P

\bibitem[\protect\citeauthoryear{{Tenorio}, {Jaffe}, {Hanany} \&
  {Lineweaver}}{{Tenorio} et~al.}{1999}]{Tenorio1999}
{Tenorio} L.,  {Jaffe} A.~H.,  {Hanany} S.,    {Lineweaver} C.~H.,  1999,
  \mnras, 310, 823

\bibitem[\protect\citeauthoryear{{Vanzella}, {Cristiani}, {Fontana}, {Nonino},
  {Arnouts}, {Giallongo}, {Grazian}, {Fasano}, {Popesso}, {Saracco} \&
  {Zaggia}}{{Vanzella} et~al.}{2004}]{Vanzella2004}
{Vanzella} E.,  {Cristiani} S.,  {Fontana} A.,  {Nonino} M.,  {Arnouts} S.,
  {Giallongo} E.,  {Grazian} A.,  {Fasano} G.,  {Popesso} P.,  {Saracco} P.,
  {Zaggia} S.,  2004, \aap, 423, 761

\bibitem[\protect\citeauthoryear{{Verde}, {Wang}, {Heavens} \&
  {Kamionkowski}}{{Verde} et~al.}{2000}]{Verde2000}
{Verde} L.,  {Wang} L.,  {Heavens} A.~F.,    {Kamionkowski} M.,  2000, \mnras,
  313, 141

\bibitem[\protect\citeauthoryear{{Vielva}, {Martinez-Gonzalez}, {Barreiro},
  {Sanz} \& {Cayon}}{{Vielva} et~al.}{2004}]{Vielva2004}
{Vielva} P.,  {Martinez-Gonzalez} E.,  {Barreiro} R.~B.,  {Sanz} J.~L.,
  {Cayon} L.,  2004, \apj, 609, 22

\bibitem[\protect\citeauthoryear{{Vielva} \& {Sanz}}{{Vielva} \&
  {Sanz}}{2010}]{Vielva2010}
{Vielva} P.,  {Sanz} J.~L.,  2010, \mnras, 404, 895

\bibitem[\protect\citeauthoryear{{Yadav} \& {Wandelt}}{{Yadav} \&
  {Wandelt}}{2010}]{Yadav2010}
{Yadav} A.~P.~S.,  {Wandelt} B.~D.,  2010, Advances in Astronomy, 2010

\end{thebibliography}

\end{document}